\documentclass[apj]{emulateapj}
\usepackage{amsmath}

\newcommand{\DDir}{\relax{D\kern-.7em{/}}}

\newcommand{\inv}[1]{\frac{1}{#1}}

\newcommand{\be}{\begin{equation}}
\newcommand{\ee}{\end{equation}}
\newcommand{\bea}{\begin{equation*}}
\newcommand{\eea}{\end{equation*}}

\newcommand{\abs}[1]{\left\vert#1\right\vert}

\newcommand{\nin}{\relax{\in\kern-.8em{/}}}

\newcommand{\al}{\alpha}

\newcommand{\om}{\omega}
\newcommand{\sig}{\sigma}

\newcommand{\vep}{\varepsilon}

\newcommand{\erg}{\mbox{ erg}}

\newcommand{\sref}{\S~\ref}

\newcommand{\jcap}{JCAP}

\begin{document}

\title{Low X-ray emission challenges supernovae remnants as the source of cosmic-ray electrons}
\author{Boaz Katz\altaffilmark{1}}
\altaffiltext{1}{Institute for Advanced Study, Einstein Drive, Princeton, New Jersey, 08540, USA;  John Bahcall Fellow}

\begin{abstract}
The X-ray synchrotron emission of each of the young supernova-remnants (SNRs) SN1006, Kepler, Tycho, RCW86 and Cas A, is roughly given by $\nu L_{\nu}\sim 10^{45}\erg/t$, where $t$ is the remnant's age. The electrons emitting the X-ray emission cool fast, implying that the X-ray emission is calorimetric and equal to half of the cosmic ray (CR) electron acceleration efficiency (per logarithmic interval of particle energies, at multi TeV energies). Assuming Sedov-Taylor expansion, the resulting CR electron yield per SNR is estimated to be $E^2dN_e/dE\approx 6\nu L_{\nu}t \sim 10^{46}\rm erg$. This is about two orders of magnitudes below the required amount for explaining the observed electron CRs at $E\sim 10\rm GeV$. Possible resolutions are 1. a soft acceleration spectrum allowing much more energy at $E\sim 10\rm GeV$ compared to $E\sim 10\rm TeV$, 2. an increased acceleration efficiency at later phases of the SNR evolution (unlikely), or 3. SNRs are not the source of CR electrons.\end{abstract}

\section{X-rays: Electron CR yield per SNR is  $\sim 10^{46}\rm erg$}\label{sec:X-rays}

One of the most exciting developments in the high energy study of supernovae remnants (SNRs) is the identification of non-thermal X-rays which are likely due to synchrotron emission of multi-TeV accelerated electrons \citep[e.g.][]{Koyama95,Fink94,Allen97}. An interesting aspect of this emission is that the non-thermal flux of the nearby young SNRs , Cas A, Kepler, Tycho, RCW86 and SN1006 is found to be of similar magnitude, $\nu f_{\nu, X}\sim 10^{-10} \rm ergs~cm^{-2} ~s^{-1}$ (see table \ref{table:SNRs}). This is in striking contrast with the non-thermal radio flux of these remnants. The radio flux of Cas A is $\sim 100$ times larger than that of other nearby remnants. 

\begin{table}[ht] 
\centering 
\caption{Nearby young, shell supernova remnants} 
\begin{tabular}{c c c c c c} 
\hline\hline 
SNR & d\footnote{Adopted from \citet{Green09} based on expansion and shock velocity estimates from proper motions or $H\alpha$ line widths.} & $t^{\rm a}$& $f_{\nu}(1\rm{GHz})^a$& $\nu f_{\nu,X}$\footnote{Non thermal emmision, estimated from \citep[][]{Berezhko06,Araya10,Giordano12,Lemoine-Goumard12,Acero10}} & $\mathbf{\nu L_{\nu,X}\cdot t}$\\
& [kpc] & [yr] & [Jy] &$[\rm erg~ cm^{-2} s^{-1}]$& [$\mathbf{10^{45}} \rm \textbf{erg}$] \\[1 ex]
\hline 
Cas A & 3.4 & $\sim 300$ & 2700 	& $\sim 3\times10^{-10}$	& $\sim \mathbf{4}$  \\ [1ex] 
Kepler& 2.9 & $400$ 		& 20 		& $\sim 3\times10^{-11}$	& $\sim \mathbf{0.4}$\\
Tycho&  2.4 & $400$		& 60		& $\sim 10^{-10}$		& $\sim \mathbf{1}$\\
RCW86&2.3& $\sim 2000 \footnote{Assuming this is the remnant of the SN at 186AD \citep[e.g.][]{Stephenson02}}$& 50		& $\sim 10^{-10}$		& $\sim \mathbf{4}$\\
SN1006&2.2& $ 1000$		& 20		& $\sim 10^{-10}$		& $\sim \mathbf{2}$\\
\hline 
\end{tabular} 
\label{table:SNRs} 
\end{table} 

The likely explanation for the significant difference between the radio and X-ray properties (both emitted by accelerated electrons which interact with the magnetic field in the remnant) is that the high energy electrons that emit the X-rays are efficiently cooled by this emission \citep[e.g.][]{Vink03}. While the radio emission strongly depends on the magnetic field value $B$, $L(\rm 1GHz)\propto B^{3/2}$, the X-ray is calorimetric and independent of $B$, being proportional to the acceleration rate. The difference in radio flux thus likely results from a large difference in magnetic field value. The condition for fast cooling can be verified directly by comparing the cooling time and the age of the remnants. The cooling time, $t_{\rm cool}=\vep/\abs{\dot\vep}$, of an isotropic distribution of electrons with energies $\vep=\gamma m_e c^2$, moving through a magnetic field $B$ and therefore emitting synchrotron with a luminosity $-\dot \vep=(4/3)\gamma^2\sig_T(B^2)/(8\pi)c$ at a typical frequency $\nu\approx \gamma^2 eB/(10 m_e c^2)$ is approximately:
\begin{align}\label{eq:tcool}
t_{\rm cool}\approx& 60\rm{yr}\left(\frac{\vep}{20 \rm TeV}\right)^{-1}\left(\frac{B}{100\mu \rm G}\right)^{-2}\cr
\approx& 60\rm{yr} \left(\frac{h\nu}{\rm keV}\right)^{-1/2}\left(\frac{B}{100\mu \rm G}\right)^{-3/2}.
\end{align}
The magnetic fields in these remnants have to be large given that the X-ray emission is $\gtrsim 100$ brighter than the TeV emission \citep{Albert07,Aharonian08,Aharonian09,Acero10,Acciari11}. A minimal magnetic field
\begin{equation}
B\gtrsim30\mu\rm{G} \left(\frac{\nu L_{\nu,\rm TeV}}{0.01\nu L_{\nu,\rm X}}\right)^{-1/2}
\end{equation}
is required so that the X-ray emission is sufficiently brighter than the Inverse Compton emission from the same electrons as they interact with the CMB photons. The large magnetic fields imply cooling times of multi-keV emitting electrons which are shorter than the age of the remnants, supporting the fast-cooling interpretation.

Given that the X-rays are calorimetric, they directly probe the acceleration efficiency of electrons,
\begin{equation}
\nu L_{\nu,\rm syn}=\frac12\vep^2\frac{d\dot N_{e,SNR}}{d\vep},
\end{equation}
where ${d\dot N_{e,SNR}}/{d\vep}$ is the generation rate of accelerated electrons at the shock and the factor of $0.5$ is due to the fact that the synchrotron frequency is proportional to the square of the electron energy so $d\log\vep=0.5d\log\nu$. We next relate this luminosity to the total yield of electrons. Like other cosmic rays, the electrons are trapped within the SNR and eventually lose their energy due to adiabatic expansion. Once the SNR becomes radiative, the CRs can escape. We conservatively assume that the CRs in the remnant at the latest phases escape the SNR without further losses. Assuming a constant fraction of the thermal energy behind the shock is converted to CRs, the electron CR energy is constant during the Sedov-Taylor phase \citep[e.g.][]{Chevalier83}. This CR electron energy, $\vep^2dN_{e,SNR}/d\vep$ (assumed to be independent of $\vep$), is given by 
\begin{equation}
\vep^2\frac{dN_{e,SNR}}{d\vep}=A_{r}\vep^2\frac{d\dot N_{e,SNR}}{d\vep}t\approx 6 \nu L_{\nu,\rm syn} t
\end{equation}
 where $A_{r}\approx 3$ is a dimensionless coefficient which is approximately equal to $3$ and is calculated in \sref{sec:ST} based on the results of \cite{Chevalier83}.      

Given the non-thermal X-rays of these SNRs, $ \nu L_{\nu,\rm syn} t\sim 1\times 10^{45}\rm erg$, the implied CR electron yield is of order $\sim 10^{46}\rm erg$, which is a tiny fraction of the supernova energy, $E_{\rm SNR}\sim 10^{51}\rm erg$. In fact, as we next argue, this is about two orders of magnitudes smaller than the required yield in order to account for the observed CR electrons.

\section{Observed yield $(10^{46}\rm erg)$ is two orders of magnitude less than the required yield $(10^{48} \rm erg)$}
Consider next the yields required in order to produce the CR electrons observed at earth. We focus on electrons with energies $\vep\sim 10\rm GeV$, which are sufficiently energetic to be unaffected by solar modulation and not energetic enough to be susceptible to energy losses within the lifetime ($~10^5\rm yr$) of a SNR (see equation \ref{eq:tcool}, with a typical late-phase magnetic field of $B\sim 10\mu \rm G$).
The production rate of electrons at $\vep\sim 10\rm GeV$ is roughly \citep[e.g.][section \sref{sec:CRelectrons}]{Strong10}.
\begin{equation}
\vep^2d\dot N_{e,\rm MW}/d\vep \sim 3\times 10^{38} \rm erg~s^{-1}.
\end{equation}
  Assuming a galactic supernova rate of $\dot N_{\rm SN, MW}\sim 1/(50~ \rm yr)$, the required output of electrons per SNR is
\begin{equation}
\vep^2\frac{dN_{e,SNR,req}}{d\vep}=\vep^2\frac{d\dot N_{e,\rm MW}}{d\vep}/\dot N_{\rm SN, MW}\sim 5\times 10^{47}\rm erg,
\end{equation}

which is $1.5-2$ orders of magnitude larger than the observed one. A few possibilities for resolving this discrepancy are:
\begin{enumerate}
\item The acceleration efficiency at low, $10 \rm GeV$ energies is much higher than the at the  $\sim 10 \rm TeV$ energies where electrons emit the synchrotron X-rays. An upper limit for the acceleration at low energies can be obtained by the radio observations:
\begin{align}
&\vep^2\frac{dN_{e,SNR}}{d\vep}=3\times 10^{46}\rm erg\cr 
&\times \left(\frac{\nu}{\rm GHz}\right)^{-1/2}\frac{f_{\nu,\rm syn}}{30 \rm Jy}\left(\frac{B}{100\mu G}\right)^{-3/2}\left(\frac{d}{3\rm kpc}\right)^2.
\end{align}
For the lowest allowed magnetic field values of $\sim 30 \mu \rm G$, an energy of $\sim 2\times 10^{47}\rm erg$ is possible, bridging a significant part of the gap.
\item The acceleration efficiency becomes stronger at later phases of the SNR evolution. We find this unlikely given that the shock slows with time implying lower thermal particle energies and likely smaller injection to the acceleration process.

%
\item SNRs are not the source of CR electrons.
\end{enumerate}

The uncertainties in these rough estimates are sufficiently large that SNRs cannot be ruled out as the source of CR electrons. However, given the likely decline in CR acceleration efficiency at later times when the shock is slower, and taking into account that some adiabatic losses are expected when the CRs are eventually released, we believe that the observed low efficiency poses a serious challenge. The most exciting possibility is that SNRs are not the main source of CR electrons and that the actual source is yet to be found. 

\acknowledgments 
We thank Kohta Murase and Kfir Blum for useful discussions.

\appendix

\section{Relation between energy and luminosity in the Sedov-Taylor phase}\label{sec:ST}
Following \citet{Chevalier83}, we consider a spherical blast wave propagating with a shock velocity $v_s=\dot r_s$ into a surrounding medium with constant density $\rho_0$ . It is assumed that a constant fraction $\om$ of the pressure in the immediate post shock region is carried by accelerated cosmic rays, $\om= p_{\rm CR}/(p_{\rm CR}+p_t)$. Assuming the cosmic rays do not diffuse significantly and are always relativistic, the hydrodynamic evolution at late times is self similar.  The radius evolves similarly to the Sedov-Taylor solution
\begin{equation}\label{eq:Et}
r_s=\left(\frac{\al E}{\rho_0}\right)^{1/5}t^{2/5}
\end{equation}
where $E$ is the total, constant energy and $\al$ a dimensionless number that depends on $\omega$. Equation \eqref{eq:Et} implies that
\begin{equation}\label{eq:vsrst}
v_s=\frac25\frac{r_s}{t}
\end{equation}

Conservation of energy, mass and momentum across the shock implies that 
 \begin{equation}\label{eq:cons}
\frac{\rho_0}{\rho_s}=\frac{\gamma_s-1}{\gamma_s+1},~~~~p_{\rm CR}=\frac{2\om}{\gamma_s+1}\rho_0v_s^2
\end{equation}
where 
\begin{equation}\label{eq:gammas}
\gamma_s=\frac{5+3\om}{3(1+\om)},
\end{equation}
and $\rho$ is the post-shock density. 

\begin{figure}[h!]
\centering
\includegraphics[width=0.5\textwidth]{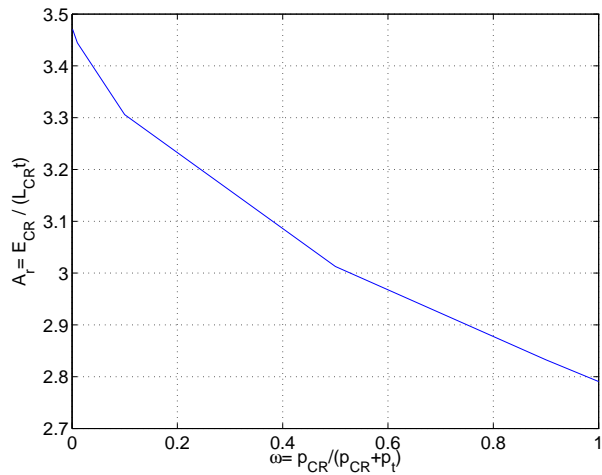}
\caption{The relation between the CR energy $E_{\rm CR}$, the age $t$ and the acceleration rate $L_{\rm CR}$ in Sedov-Taylor expansion based on the results of \citet{Chevalier83}. The energy is given by $E_{\rm CR}=A_r L_{\rm CR}t$, where $A_r$ is a function of the acceleration efficiency, $\omega=p_{\rm CR}/(p_{\rm CR}+p_t)$. $A_r$ is calculated using equation \eqref{eq:ELt}, and the values of $\om$,$\al$,and $E_{\rm CR}/E$ obtained by \citet[][table 6]{Chevalier83}.
\label{fig:Ar}}
\end{figure}

Using equations \eqref{eq:cons},\eqref{eq:vsrst}, the CR acceleration rate can be expressed as
\begin{equation}\label{eq:Lr}
L_r=\dot M\frac{3p_{\rm CR}}{\rho_s}=\frac{192\pi\om(\gamma_s-1)}{125(\gamma_s+1)^2}\frac{\rho_0r_s^5}{t^3}
\end{equation}
where
\begin{equation}
\dot M_s=4\pi r_s^2v_s\rho_0
\end{equation}
is the mass flux across the shock and $3p_{\rm CR}/\rho_s$ is the energy carried by CRs per unit mass in the immediate post-shock region.
Using equations \eqref{eq:Et},\eqref{eq:Lr} the following relation between the total energy in CRs and the CR acceleration rate is obtained,
\begin{equation}\label{eq:ELt}
A_r=\frac{E_r}{L_rt}=\frac{E_r}{E}\frac{E}{L_rt}=\frac{E_r}{E}\frac{125(\gamma_s+1)^2}{192\pi\om(\gamma_s-1)\al}.
\end{equation}
For $\om=0.01,0.1,0.5,0.9,0.99$ \citet{Chevalier83} found the values $\al=1.98,1.70,1.18,1.00,0.976$ and $E_r/E=0.031,0.24,0.64,0.78,0.80$ respectively. Using equations \eqref{eq:ELt},\eqref{eq:gammas} we find for all $\om$ that (see figure \ref{fig:Ar})
 \begin{equation}
 A_r=3\pm0.5.
 \end{equation} 
This result can be applied to a subset of accelerated CRs and in particular to the electrons accelerated to a given logarithmic interval of energies. It should be kept in mind that due to adiabatic expansion, the electrons estimated in this way will have different energies at different locations in the remnant, but with a constant logarithmic interval.  
 
\section{CR electron luminosity of the Milkey Way}\label{sec:CRelectrons}

Given that positrons and electrons have similar trajectories and energy losses as they propagate through the galaxy, the production rate of electrons can be calculated using the measured electron to positron ratio and the known production rate of positrons:
\begin{align}
\vep^2\frac{d\dot N_{e,MW}}{d\vep}\sim  \frac{n_{e^-}}{n_e^{+}}\vep^2\frac{d\dot N_{e^+}}{d\vep}\sim 4\times 10^{38}\rm erg s^{-1}~~~~~(\rm{at}~\vep\sim 10\rm GeV)
\end{align}
where the positron fraction at $10\rm GeV$ is measured to be \citep{Adriani09,Aguilar13}
\begin{equation}
\frac{n_e^{-}}{n_{e^+}}\approx 20
\end{equation}
and the production rate of positrons is given by 
\begin{equation}
\vep^2\frac{d\dot N_{e^+}}{d\vep}\approx \inv{\rho_{\rm ISM}}\vep^2\frac{d\dot n_{e^+}}{d\vep}M_{\rm gas}\sim 2\times 10^{37}\rm erg s^{-1}
\end{equation}
where $M_{\rm gas}\sim 10^{10} M_{\rm sun}$ is the gas mass in the galaxy, and the production rate of positrons per unit ISM mass is \citep[e.g.][]{Katz10, Blum13}:
\begin{equation}
\inv{\rho_{\rm ISM}}\vep \frac{d\dot n_{e^+}}{d\vep}|_{10\rm GeV}=\zeta_{A>1}C \left(\vep\frac{dn_p}{d\vep}\right)|_{100\rm GeV}\frac{\sig_{pp,0}}{m_p}c
\end{equation}
where $C=0.6$ is found from $p-p$ cross sections, $\zeta_{A>1}\approx 1$ is the correction due to the presence of CRs other than protons, $\sig_{pp,0}\equiv 30\rm mb$ and  $\vep\frac{dn_p}{d\vep}(10\rm {GeV})\approx 2.1\times 10^{-13} \rm {cm^{-3}} $ is the proton CR density \citep[per logarithmic particle energy,][]{Moskalenko98}, implying 
\begin{equation}
\inv{\rho_{\rm ISM}}\vep^2\frac{d\dot n_{e^+}}{d\vep}\approx 1.1\times 10^{-6}\rm erg~s^{-1}~g^{-1}.
\end{equation}

%


\bibliographystyle{apj}

\end{document}